\documentclass[10pt,aps,pra,twocolumn]{revtex4-2}
\usepackage{amsthm}
\usepackage{graphicx}
\usepackage{dcolumn}
\usepackage{bm}
\usepackage{amsmath}
\usepackage{physics}
\usepackage{xcolor}

\DeclareMathOperator{\arcsinh}{arcsinh}

\DeclareMathOperator{\arccoth}{arccoth}

\DeclareMathOperator{\diag}{diag}

\newcommand{\mathbbm}[1]{\text{\usefont{U}{bbm}{m}{n}#1}}
\usepackage{subfig}

\usepackage{ragged2e}
\newtheorem{conjecture}{Conjecture}
\newtheorem{condition}{Condition}

\usepackage[compatibility=false]{caption}
\usepackage{indentfirst}
\usepackage{titlesec}
\usepackage{caption}

\titlespacing*{\section}{0pt}{*1.2}{*0.8}       
\titlespacing*{\subsection}{0pt}{*1.0}{*0.6}

\begin{document}

\preprint{APS/123-QED}

\title{Quantum-Enhanced Change Detection and Joint Communication-Detection}

\author{ Zihao Gong$^{(1)}$ and Saikat Guha$^{(1,2)}$}
\affiliation{$^{(1)}$Department of Electrical and Computer Engineering, University of Maryland, College Park MD}
\affiliation{$^{(2)}$Wyant College of Optical Sciences, University of Arizona, Tucson AZ}

\begin{abstract}
Quick detection of transmittance changes in an optical channel is crucial for secure communication. We demonstrate that preshared entanglement using two-mode squeezed vacuum states significantly reduces detection latency compared to classical and entanglement-augmented coherent-state probes. The change-detection latency is inversely proportional to the quantum relative entropy (QRE), which goes to infinity in the absence of thermal noise, suggesting idealized instantaneous detection. However, in realistic scenarios, we show that QRE scales logarithmically with the inverse of the thermal-noise mean photon number. We propose a receiver that achieves this scaling and quantify its performance gains over existing methods. Additionally, we explore the fundamental trade-off between communication capacity and change-detection latency, highlighting how preshared entanglement enhances both.

\end{abstract}

\maketitle


\section{Introduction} \label{sec:introduction}
Quick detection of a change in channel transmittance plays a vital role in safeguarding optical network integrity. These changes can stem from either malicious tapping or environmental variations. One of the most effective classical approaches for real-time change detection is the cumulative sum (CUSUM) test \cite{Page1954}, which continuously evaluates statistical deviations, signaling an event when accumulated metrics surpass predetermined false-alarm rate. The CUSUM test minimizes the worst-case average detection delay subject to a constraint on the false-alarm rate and is thus optimal under Lorden's formulation of quickest change detection\cite{Moustakides1986,pollak1987average,lorden1971procedures}.

Quantum-enhanced techniques offer novel opportunities for improving change detection performance. 
While the ultimate performance of classical systems is governed by the relative entropy between post- and prechange probability distribution, the limit in quantum system is determined by the quantum relative entropy (QRE) between post- and prechange quantum states \cite{fanizza2023ultimate,john2025fundamentallimitsquickestchangepoint}. When constrained to a specific receiver, the quantum problem reduces to its classical analog. The worst-case minimum change-detection latency  scales inversely to the relative entropy~\cite{lai1998information}. Attaining QRE may require joint measurement, which introduces additional latency, yet maximizing the QRE remains paramount for achieving the quickest possible change detection.

\begin{figure}[ht] 
    \centering
       \includegraphics[width=0.85\linewidth]{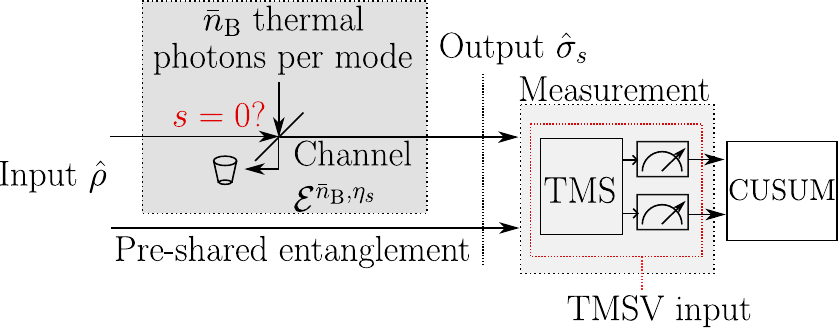}
  \caption{\justifying 
In a lossy thermal noise bosonic channel $ \mathcal{E}^{\bar{n}_{\rm B}, \eta_s} $, detecting abrupt transmittance changes from $ \eta_0 $ to $ \eta_1 $ at an unknown time $ \tau_c $ is important. A quantum state $ \hat{\rho} $ is continuously sent through the channel, with the potential to utilize pre-established entanglement with the receiver. The output state $ \hat{\sigma}_s $ is then subject to measurement, and the collected data are processed using a CUSUM test. The receiver setup, indicated by a red square in the measurement block, includes a two-mode squeezing (TMS) operation followed by one or two photon-number-resolving (PNR) detector or other appropriate measurement configurations when the input state $ \hat{\rho} $ is a TMSV state.  }
  \label{fig:channel} 
\end{figure}

Our work focuses on entanglement-enhanced CUSUM testing for tap detection in a lossy thermal-noise bosonic channel $\mathcal{E}^{\Bar{n}_{\rm B},\eta_s}$,  as illustrated in Fig.~\ref{fig:channel}. This channel describes a single-mode electromagnetic-field transmission experiencing linear loss and additive thermal noise. It is modeled by a beam splitter with transmittance $\eta_s$ ($s = 0,1$), with the environment injecting a thermal state with mean photon number $\Bar{n}_{\rm B}$. Initially, the transmittance is $\eta_0$. At an unknown (discrete) time $t = n_c$, an adversary taps the channel, which introduces an additional loss and reduces the transmittance to $\eta_1 = \eta_0\eta_{\rm tap}$. The transmitter encodes quantum states $\hat{\rho}$ into each channel use and potentially leverages preshared entanglement with the receiver. The input state $\hat{\rho}$ is constrained by a mean photon number per mode $\Bar{n}$. The receiver measures the output state $ \hat{\sigma}_{s}$, applies the CUSUM test to the data, and declares a change at some time $t = n_d$. The receiver's goal is to maximize the relative entropy to minimize the expected latency $t= n_d - n_c$.

Quantum probes exploiting squeezing and entanglement have been shown to enhance precision in transmittance sensing \cite{zihao2023,nair20qi,LloydQuantumIlluminination2008}, and prior work has established that preshared entanglement can increase the communication capacity of quantum channels \cite{saikat2020ISIT}. 
Motivated by these advantages, we propose an entanglement-enhanced strategy for detecting transmittance  changes in  $\mathcal{E}^{\bar{n}_{\rm B },\eta_s}$. In particular, we analyze the performance of a transceiver based on a two-mode squeezed vacuum (TMSV) state and identify a receiver design that asymptotically achieves the fundamental QRE bound.

Although quantum change detection and quantum communication have been studied independently, their integration remains largely unexplored. Existing work on quantum change detection assumes that the receiver has perfect knowledge of the transmitted codeword \cite{Xiong2023ISIT}. In realistic scenarios, however, the receiver must jointly decode an unknown message and simultaneously detect changes in the channel conditions. We develop a unified quantum framework that enables simultaneous communication and change detection by applying a CUSUM test to a mixture of codeword-induced output distributions. By jointly optimizing the transmitter and receiver design, we explore the trade-off between communication capacity and change-detection delay, aiming to enhance the overall system performance.

The rest of this paper is organized as follows. In Sec.~\ref{sec:QRE of TMSV and coh}, we derive the QRE of TMSV and coherent states, emphasizing their scaling behavior with noise in the low-noise regime. We also introduce the receiver that achieves the infinite relative entropy in a pure-loss channel. In Sec.~\ref{sec:numerical and comparison}, we analyze the performance of different transceivers. 
While achieving the QRE may require joint measurement, which introduces additional latency, we demonstrate that a 
two-mode squeezer (TMS) followed by a photon-number-resolving (PNR) detector asymptotically attains the QRE in the limit of an infinite input mean photon number $\bar{n}\to \infty$ without the need for joint measurement. Furthermore, we propose a conjecture that the TMSV state is optimal among two-mode Gaussian states and present numerical evidence as well as a theoretical analysis.
 In Sec.~\ref{sec:joint transmittance}, we present the joint communication and change detection problem, showing that preshared entanglement simultaneously enhances channel capacity and reduces change detection latency. We also numerically simulate the CUSUM test to illustrate the detection process and demonstrate the superior performance of the binary phase-shift keying (BPSK) modulated TMSV state.

\section{Quantum relative entropy of TMSV and coherent states} \label{sec:QRE of TMSV and coh}

We apply the formula for QRE between $ m $-mode Gaussian states from Eq.~(9) of \cite{PhysRevLett.119.120501}. The QRE expressions for a TMSV input state, $ D_{\rm TMSV} $, and a coherent-state input, $ D_{\rm coh} $, are derived in Appendix \ref{app:QRE}.  

This section examines the low-noise regime where $ \bar{n}_{\rm B} \to 0 $, reducing the lossy thermal-noise channel $ \mathcal{E}^{\bar{n}_{\rm B},\eta_s} $ to a pure-loss channel in this limit. Under these conditions, the QREs for both the coherent state ($ D_{\rm coh} $) and the TMSV state ($ D_{\rm TMSV} $) diverge to infinite, suggesting the theoretical possibility of instantaneous change detection. However, this is practically unattainable because both $ D_{\rm coh} $ and $ D_{\rm TMSV} $ scale as $ -\ln(\bar{n}_{\rm B}) $ when $ \bar{n}_{\rm B} \to 0 $. The specific scaling behaviors are given by  

\begin{align}
    \lim_{\bar{n}_{\rm B}\to 0 } -\frac{D_{\rm coh}}{\ln(\bar{n}_{\rm B})} &= \bar{n}\left(\sqrt{\eta_0} - \sqrt{\eta_1}\right)^2, \\
    \lim_{\bar{n}_{\rm B}\to 0 } -\frac{D_{\rm TMSV}}{\ln(\bar{n}_{\rm B})} &= \frac{\bar{n}(1+\bar{n})\left(\sqrt{\eta_0} - \sqrt{\eta_1}\right)^2}{1+\bar{n}(1+\eta_0)}.
\end{align}  
Furthermore, $ D_{\rm TMSV} $ diverges strictly faster than $ D_{\rm coh} $, as described by  
\begin{align}
   \lim_{\bar{n}_{\rm B}\to 0 } \frac{D_{\rm TMSV}}{D_{\rm coh}} = \frac{1+\bar{n}}{1+\bar{n}(1-\eta_0)}>1.
\end{align}  
This result indicates the advantage of the TMSV state over the coherent state in the low-$ \bar{n}_{\rm B} $ regime \cite{fanizza2023ultimate}.  

For the coherent state, infinite QRE can be realized using a Kennedy receiver \cite{kennedy1973near}. This approach involves applying a displacement operator $ \hat{D}(-\sqrt{\eta_0\bar{n}}) $, followed by single-mode photon detection. The displacement operation converts the prechange state into a vacuum state, while the postchange state remains nonvacuum, ensuring that photons are detected only in the postchange case.  

Similarly, achieving infinite QRE is theoretically achievable for the TMSV state. When $ \bar{n}_{\rm B} = 0 $, the output state $ \hat{\sigma}_{\mathrm{T},s} $ is given by  $ \hat{\sigma}_{\mathrm{T},s} = \hat{S}^{\dagger}(r_s)\left(\ket{0}\bra{0}\otimes \hat{\sigma}_{\rm th}\right)\hat{S}(r_s) $,
where $ \hat{S}(r_s) $ represents the TMS operator. If we apply an inverse TMS operation $ \hat{S}(-r_0) $, followed by single-photon detection (SPD) in the signal mode while tracing out the idler mode, an infinite relative entropy is achieved, as photon detection is exclusive to the postchange scenario.

Interestingly, for both coherent and TMSV input states, the standard CUSUM test can detect changes without requiring precise knowledge of the output distribution. The test signals a change whenever a photon is detected.  
Despite these theoretical insights,  practical challenges limit the feasibility of these results. High-precision displacement and TMS operations remain technically demanding in contemporary experimental setups.

\section{Comparison of transceivers}\label{sec:numerical and comparison}
\begin{figure*}[ht] 
    \centering
  \subfloat[$\bar{n}=5, \eta_0 = 0.9, \eta_1 = 0.8$  \label{fig:a}]{
  \null\hfill
    \includegraphics[width=0.43\linewidth]{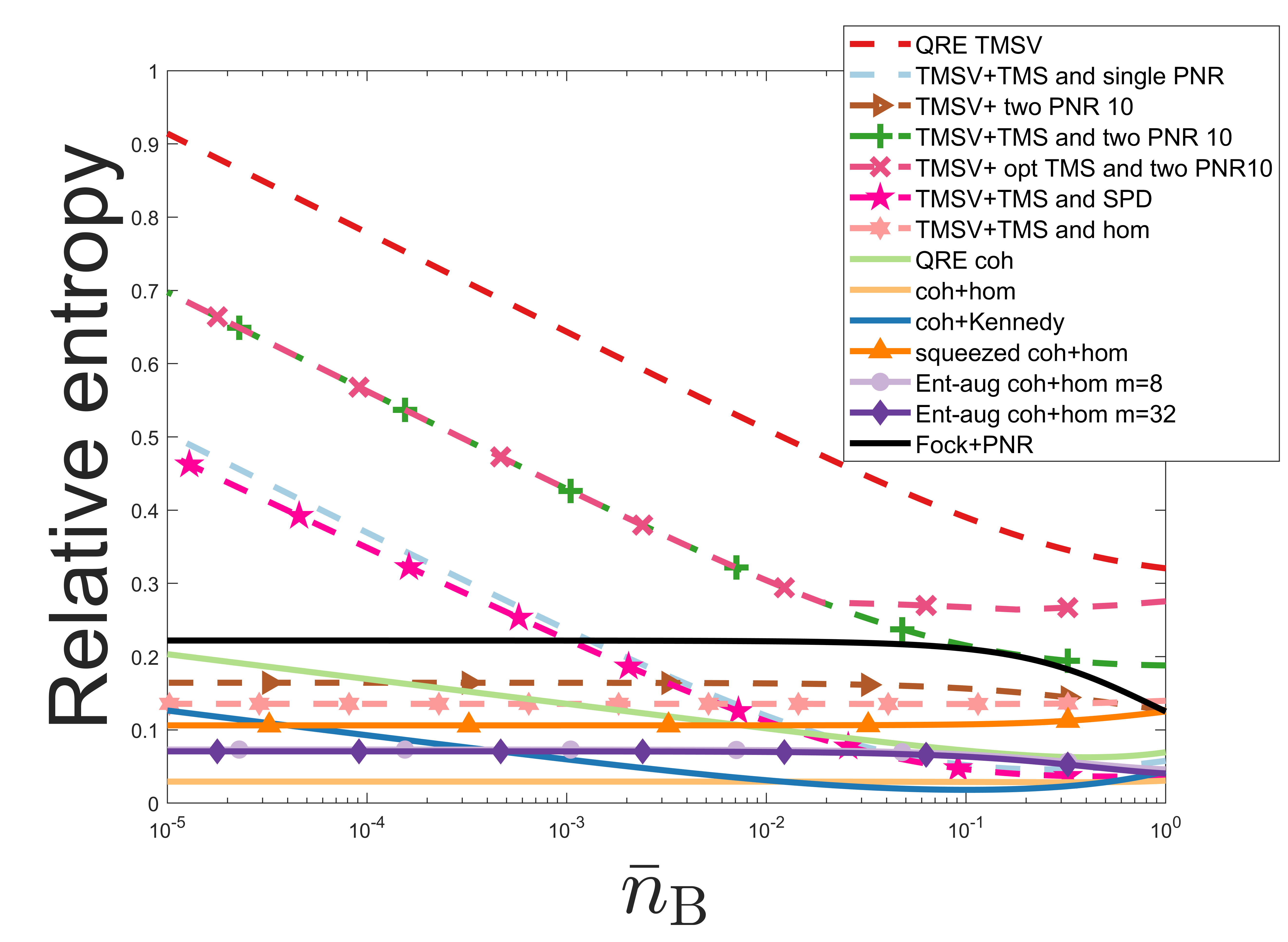}}
    \hfill
  \subfloat[$\bar{n}=400, \eta_0 = 0.9, \eta_1 = 0.8$\label{fig:b}]{
        \includegraphics[width=0.46\linewidth]{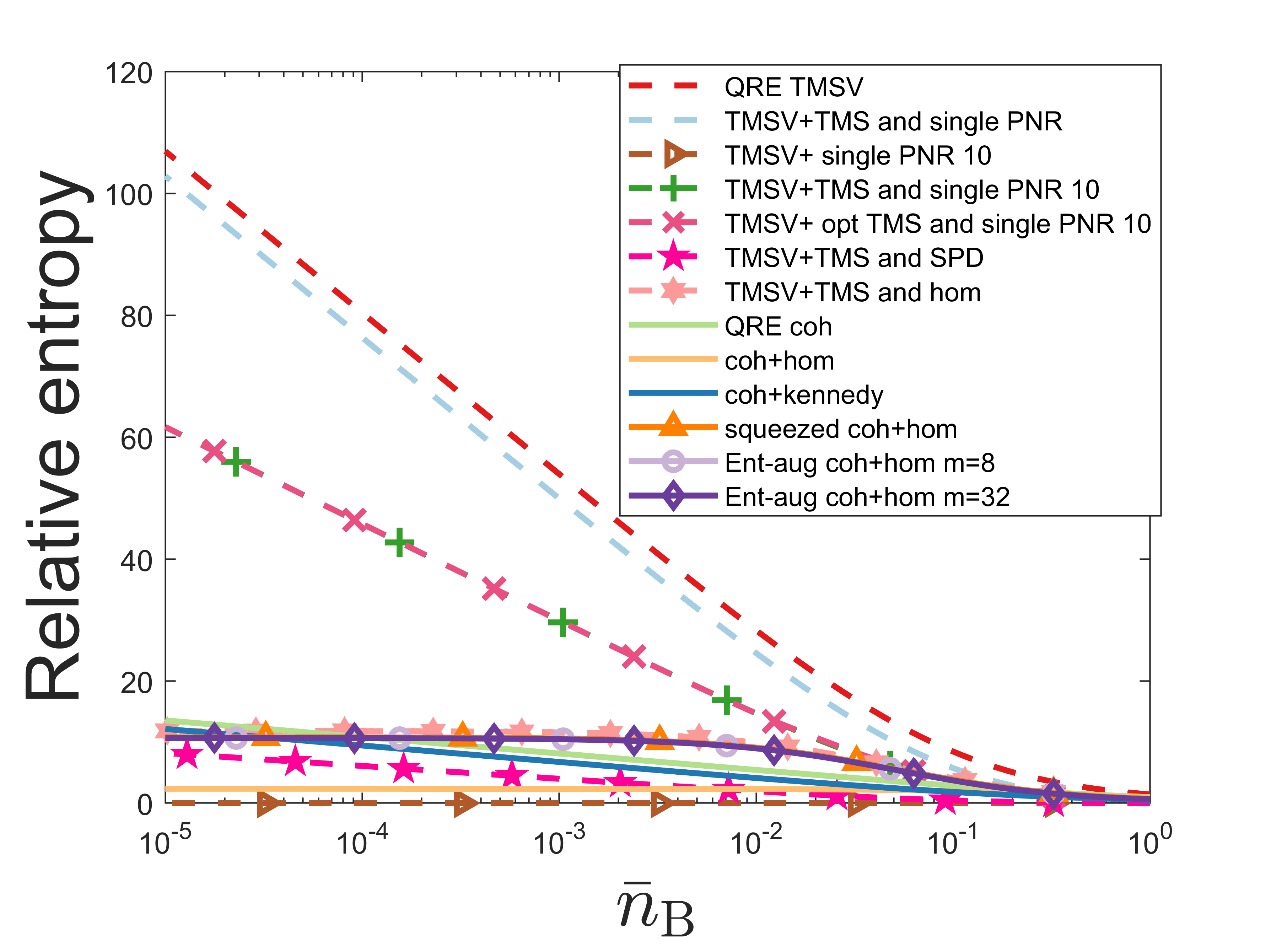}}
    \hfill\null
  \caption{\justifying Relative entropy vs mean photon number of the thermal noise level for various transceivers. The dashed curves are for the TMSV input state, and the solid curves are for the coherent input state. The number following "PNR" represents the maximum resolvable photon number of the detector, while the absence of a number indicates a ideal PNR detector. (a) $\bar{n} = 5$, $\eta_0 = 0.9$, $\eta_1 = 0.8$ (b) $\bar{n} = 400$, $\eta_0 = 0.9$, $\eta_1 = 0.8$. }
  \label{fig:RE_plot} 
\end{figure*}

We analyze different pairings of input states and receiver architectures. The input states considered include coherent, squeezed coherent, entanglement-augmented (EA) coherent~\cite{SaikatEAcoh}, and TMSV states. The receiver includes homodyne, Kennedy, and TMS followed by PNR detection. To denote the relative entropy $S$, we adopt a notation in which the first subscript specifies the input state and the second subscript indicates the receiver. We omit "TMS" when referring to PNR-based receivers. For details see Appendix \ref{sec:RE}

\subsection{Asymptotic attainment of QRE for the TMSV state}

We demonstrate that the relative entropy for the TMSV state, when measured using a TMS followed by a PNR detector on the signal mode, asymptotically approaches the QRE for the TMSV input state:
\begin{align}
    \lim_{\bar{n}\to\infty} \frac{S_{\rm TMSV,PNR}^{(\infty)}}{D_{\rm TMSV}} = 1,
\end{align}
where $ S_{\rm TMSV,PNR}^{(\infty)} $ is given in \eqref{eq:RE TMSV PNR} and the superscript $(\infty)$ indicates an ideal PNR detector with infinite photon resolution. Specifically, both $ S_{\rm TMSV,PNR}^{(\infty)} $ and $ D_{\rm TMSV} $ scale linearly with $ \bar{n} $ as $ \bar{n} \to \infty $:
\begin{align}
\lim_{\bar{n}\to\infty} \frac{S_{\rm TMSV,PNR}^{(\infty)}}{\bar{n}} &=  \lim_{\bar{n}\to\infty} \frac{D_{\rm TMSV}}{\bar{n}}\\ &=\frac{(\sqrt{\eta_1} - \sqrt{\eta_0})^2}{1-\eta_0}\ln\left(1+\frac{1}{\bar{n}_{\rm B}}\right).
\end{align}

\subsection{Numerical results: Performance comparison of transceivers}

We numerically evaluate the QRE and relative entropy for the transceivers discussed in Sec.~\ref{sec:QRE of TMSV and coh} and Appendix \ref{sec:RE}. The results are plotted in Fig.~\ref{fig:RE_plot} as a function of $ \bar{n}_{\rm B} \in [10^{-5},1] $, with $ \eta_0 = 0.9 $ and $ \eta_1 = 0.8 $. In Fig.~\ref{fig:RE_plot}, the number following "PNR" indicates the maximum photon resolution of the PNR detector, and the absence of a number represents an ideal PNR detector. For the EA coherent transceiver from Ref.~\cite{SaikatEAcoh}, the relative entropy values $ S_{\rm ec,hom}(m) $ are maximized over the displacement energy $ \alpha_r $, and the optimized values are plotted. 
The relative entropy for a Fock state with a PNR detector is omitted in Fig.~\ref{fig:b} because its probability mass function (PMF) involves hypergeometric functions, which are computationally challenging to evaluate for large $ \bar{n} $.

\textit{TMSV-state input.} In the low-noise regime, a preshared TMSV state combined with TMS and SPD on the signal mode at the receiver surpasses the performance of the coherent transceiver. Increasing the photon resolution of the PNR detector enhances the relative entropy. However, the TMSV state with TMS and homodyne detection yields suboptimal performance even when the squeezing parameter is optimized.

In Fig.~\ref{fig:a} and Fig.~\ref{fig:b}, we compare our calculated squeezing parameter with the optimized squeezing parameter, as well as the case when the squeezing parameter is zero (direct photon detection). In Fig.~\ref{fig:a}, when $ \bar{n} $ is relatively small, direct photon detection shows comparable performance between the two cases. However, in Fig.~\ref{fig:b}, when $ \bar{n} $ is large, the performance without squeezing is poor. Moreover, since both $ \bar{n}$ and $\bar{n}_{\rm B}$ contribute to the relative entropy, when  $  \bar{n} $ dominates $ \bar{n}_{\rm B} $, our calculated squeezing parameter closely matches the optimal value. However, when $\bar{n}_{\rm B} $ dominates, TMS should enhance the impact of $ \bar{n}_{\rm B} $, so an optimized TMS outperforms the TMS we calculate.

\textit{Coherent-state-related input.} The Kennedy receiver approach achieves the QRE for the coherent state when $ \bar{n} $ is large. The EA and squeezed coherent transceivers outperform the coherent transceiver in terms of relative entropy per mode, as shown in Fig.~\ref{fig:RE_plot}. However, for the EA coherent transceiver, the CUSUM test detects a change only after receiving a full block of $ m $ states, introducing additional delays. Larger values of $ m $ result in greater delays, so it is optimal to choose a smaller $ m $ when $ S_{\rm ea, coh}(m) $ are equal for different values of $ m $, as seen in Fig.~\ref{fig:b}.

\textit{Fock-state input.} While the Fock state is optimal for transmittance sensing in a pure-loss channel because it achieves the quantum Fisher information \cite{weedbrook12gaussianQIrmp}, it becomes suboptimal for transmittance change detection because $ S_{\rm Fock,PNR} $ remains finite as $ \bar{n}_{\rm B} \to 0 $, unlike TMSV and coherent state.

\begin{figure}[ht] 
    \centering
        \includegraphics[width=0.9\linewidth]{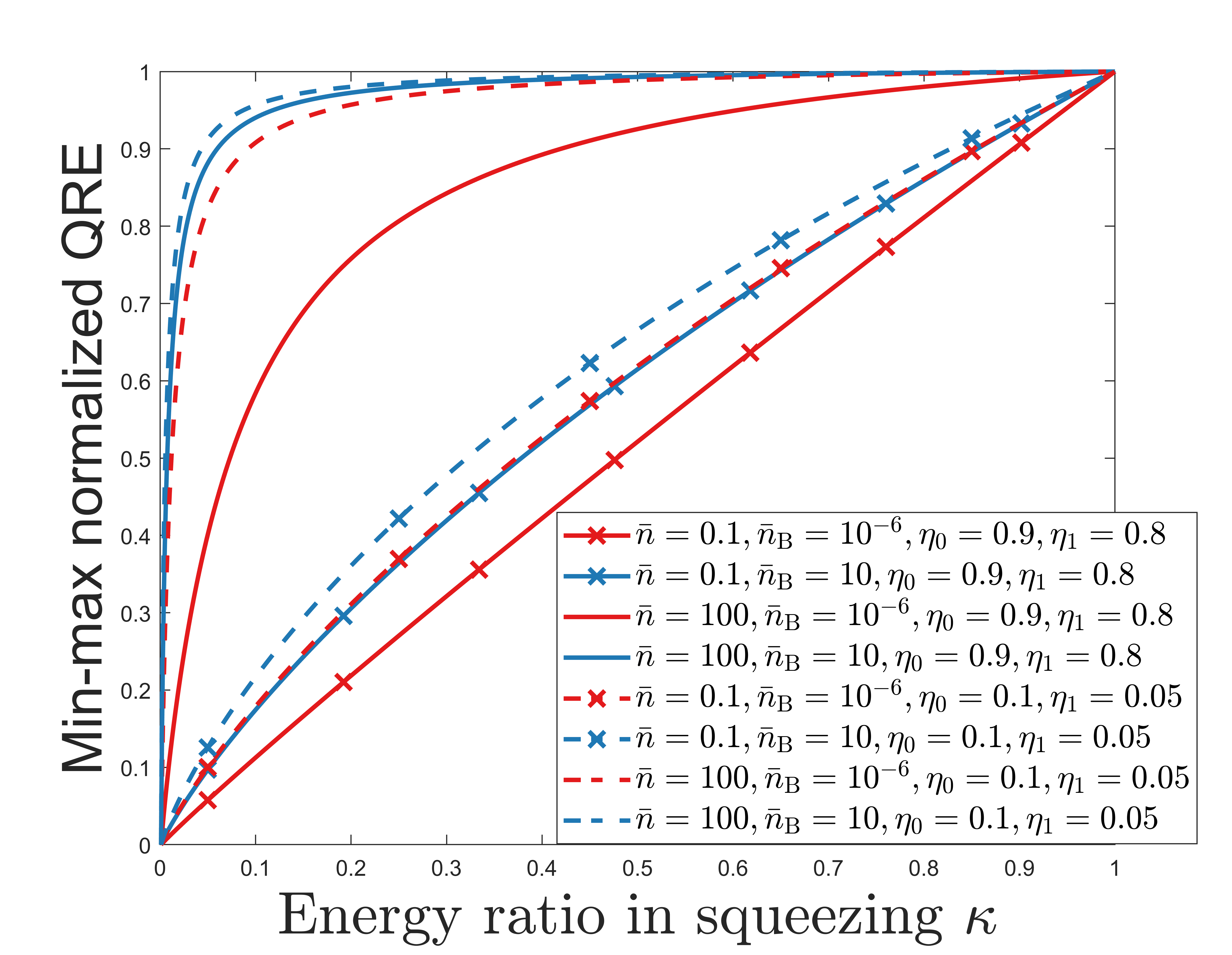}
  \caption{\justifying Proportion of energy in squeezing versus min-max normalized QRE. The solid curves are for $\eta_0 = 0.9$ and $\eta_1 = 0.8$, and the dashed curves are for $\eta_0 = 0.1$ and $\eta_1 = 0.05$. The blue curves are for $\bar{n}_{\rm B} = 10$, and the red curves are for $\bar{n}_{\rm B} = 10^{-6}$. The curves with crosses are for $\bar{n} = 0.1$, while the unmarked curves correspond to $\bar{n} = 100$.}\label{fig:Numerical evidence}
\end{figure}


\subsection{Conjecture on the optimality of the TMSV state}
In this section, we begin by formally stating a conjecture on the optimality of the TMSV state in transmittance change detection. We then identify an important intermediate condition. Finally, we show that the conjecture follows from this intermediate condition. We propose the following conjecture.

\begin{conjecture} \label{conjecture 1}
    Among all two-mode Gaussian states with a fixed photon number, the TMSV state maximizes QRE between post- and prechange states for detecting transmittance change in $\mathcal{E}^{\bar{n}_{\rm B}, \eta_s}$.
\end{conjecture}
The following condition is sufficient to conclude Conjecture \ref{conjecture 1}.
\begin{condition}\label{statement}
Let $\hat{\rho}_{\Sigma, \mu}$ denote a displaced TMSV state characterized by covariance matrix $\Sigma$ and mean vector $\mu$, with total mean photon number $\bar{n}$. Define $\bar{n}_{\Sigma}$ as the energy associated with squeezing and let $\kappa = \bar{n}_{\Sigma} / \bar{n}$ denote the energy fraction allocated to squeezing. Then, for channel  $\mathcal{E}^{\bar{n}_{\rm B}, \eta_s}$, the QRE $D(\hat{\rho}_{\Sigma, \mu})$ is maximized when $\kappa = 1$, i.e., when all energy is devoted to squeezing.
\end{condition}
We are currently unable to rigorously  prove Condition \ref{statement} due to the complexity of the QRE expression. However, we numerically evaluate $\partial_{\kappa} D\left(\hat{\rho}_{\Sigma,\mu }\right)$ across a broad range of parameter regimes, varying $\bar{n}_{\rm B}$, $\bar{n}$, $\eta_0$, $\eta_1$, and $\kappa$. In all tested cases, the partial derivative remains positive, indicating that  $D\left(\hat{\rho}_{\Sigma,\mu }\right)$ increases monotonically with $\kappa$. Furthermore, Fig.~\ref{fig:Numerical evidence} shows the min-max normalized QRE versus $\kappa$ and further supports the condition.

Now we prove Conjecture~\ref{conjecture 1} assuming the above condition holds. Consider an arbitrary two-mode Gaussian state $\hat{\sigma}_{\Sigma,\mu }$ with energy constraint $\bar{n}$. The QRE between the post- and prechange quantum state for such input satisfies the following upper bound:
\begin{align}
        D\left(\hat{\sigma}_{\Sigma,\mu } \right) &= D\left( \int p_G(\Sigma - \Sigma_0,\xi) \hat{\rho}_{\Sigma_0,\mu - \xi } d\xi \right) \label{eq:1 1}\\
    & \le  \int p_G(\Sigma - \Sigma_0,\xi) D\left( \hat{\rho}_{\Sigma_0,\mu - \xi }  \right) d\xi \label{eq:1 2} \\
    & \le D\left( \hat{\rho}_{\Sigma_0 ,\mu^\prime}  \right)\int p_G(\Sigma - \Sigma_0,\xi) d\xi \label{eq:upper 1} \\
    &\le  D\left( \hat{\rho}_{\Sigma^{\prime},0 }  \right).\label{eq:upper 2}
\end{align}
Equation \eqref{eq:1 1} comes from the fact that any Gaussian state with energy constraint $\bar{n}$ can be expressed as a Gaussian mixture of pure Gaussian states~(see Eq.~(3.32) in \cite{matsubara2019optimal}), where $\Sigma_0$ satisfies $\Sigma - \Sigma_0\ge0$ and $\hat{\rho}_{\Sigma_0,\mu - \xi }$ is a pure two-mode Gaussian state that is essentially a displacement TMSV state.  Equation \eqref{eq:1 2} is from the joint convexity of QRE. In \eqref{eq:upper 1}, $\mu^{\prime}$ is chosen such that the energy  of $\hat{\rho}_{\Sigma_0,\mu^{\prime}}$  equals $\bar{n}$. Thus, \eqref{eq:upper 1} holds since the QRE increases monotonically with displacement energy. Moreover, since  $\hat{\rho}_{\Sigma^{\prime},0}$  is a TMSV state with mean photon number  $\bar{n}$, \eqref{eq:upper 2} follows from Condition \ref{statement}. Therefore, the QRE for any two-mode Gaussian input state is upper bounded by that of a TMSV input state.

\begin{figure}[ht] 
    \centering
        \includegraphics[width=0.9\linewidth]{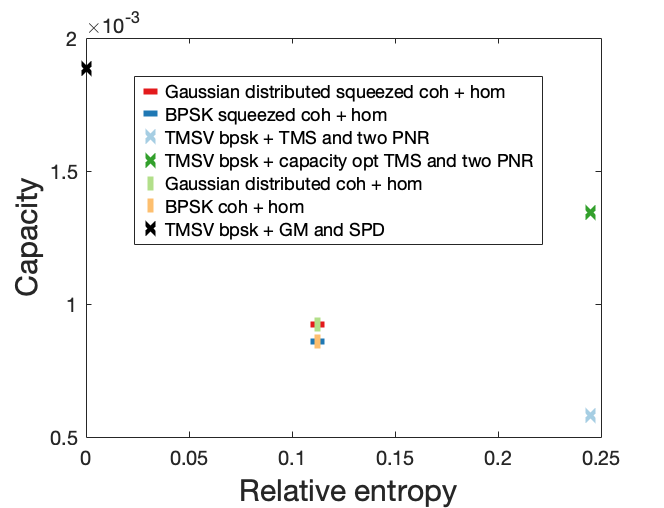}
  \caption{\justifying Relative entropy versus channel capacity in joint communication and change detection problem. $\bar{n} = 0.001 $, $\bar{n}_{\rm B} = 10$, $\eta_0 = 0.9$, and $\eta_1 = 0.8$. Alice sends a codeword to Bob, who employs CUSUM test for change detection of transmittance and decode the codeword for communication.   }\label{fig:RE vs Capacity}
\end{figure}

\section{Joint transmittance change detection and communication} \label{sec:joint transmittance}
A recent line of research explored scenarios in which Alice transmits codewords to Bob, who must simultaneously decode the messages for communication and perform a CUSUM test for change detection~\cite{SaikatEAcoh}. The codeword $ x $ is drawn from a codebook $\mathcal{C}$, and the communication channel is modeled as a discrete memoryless channel characterized by the conditional probability distribution $ p_s(Y|X) $, where $s = 0,1$. Recent work~\cite{Xiong2023ISIT} assumed that Bob has knowledge of the transmitted codeword when conducting the CUSUM test for change detection, allowing the CUSUM test to be applied directly to the conditional distribution $ p_s(Y|X) $, while the communication task still operates on $ p_s(Y) $.

In this work, we address a more realistic setting in which the receiver lacks knowledge of the codeword during the change-detection task. Consequently, the CUSUM test must operate on the mixture of probability distributions corresponding to all possible codewords, i.e., $ p_s(Y) = \int_{x\in X} p(x) p_s(Y|x)dx $. Our objective is to jointly optimize the receiver design to balance the trade-off between maximizing channel capacity for communication and minimizing latency for change detection.  
\begin{figure}[ht] 
    \centering
        \includegraphics[width=0.9\linewidth]{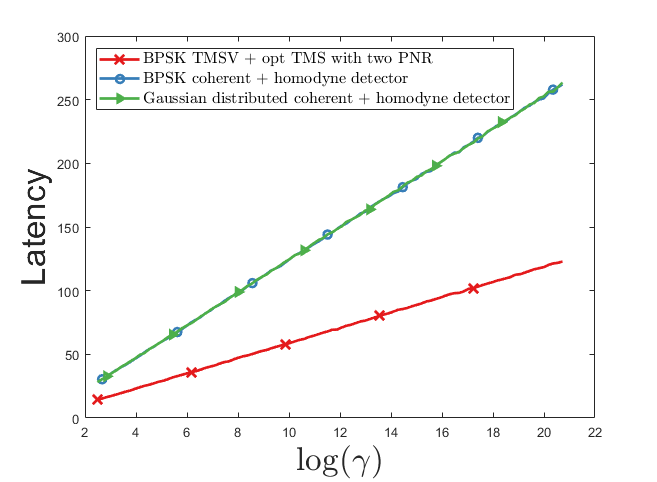}
  \caption{\justifying We numerically simulate the performance of the CUSUM test to evaluate change detection latency. Recall that the expected change-detection latency scales as $ \frac{\ln(\gamma)}{S(p_1||p_0)} $, where $\gamma$ is the threshold parameter controlling the false alarm rate and $S(p_1||p_0)$ is the relative entropy. We adopt the same parameter settings as in 
  Figure~\ref{fig:RE vs Capacity}: $\bar{n} = 0.001 $, $  \bar{n}_{\rm B} = 10 $, $ \eta_0 = 0.9 $, and $\eta_1 = 0.8$. Each plotted curve represents the average latency over 20000 independent Monte Carlo runs. }\label{fig:CUSUM_simulation}
\end{figure}

\textit{Squeezed coherent-state transceiver.} For the lossy thermal-noise bosonic channel, we analyze squeezed coherent input state $ \ket{x; r} $ with BPSK or Gaussian modulation. In the BPSK case, the modulation variable $ x $ takes a value of $ -\alpha $ or $ \alpha $, satisfying the energy constraint $ \alpha^2 + \sinh^2 r = \bar{n} $. For Gaussian modulation, $ x \sim \mathcal{N}(0, \sigma^2) $, with the total mean photon number satisfying $ \sigma^2 + \sinh^2 r = \bar{n} $. The received signal is detected using a homodyne detector in both scenarios.  

In 
Figure~\ref{fig:RE vs Capacity}, we compare the cases where $ r = 0 $,  such that the input is a coherent state and where $ r $ is optimized to maximize channel capacity. The results indicate that in the regimes of low input power, high noise, and high transmittance, allocating all energy to displacement (i.e., setting $ r = 0 $) is preferable.

\textit{BPSK TMSV-state transceiver.} We propose using a BPSK-modulated preshared TMSV state, whose covariance matrix is given by  
\begin{align}
\Sigma_{\hat{\rho}_i}=\left[\begin{array}{cccc}
\mu_1 & (-1)^{i}\mu_2 & 0 & 0 \\
(-1)^{i}\mu_2& \mu_1 & 0 & 0 \\
0 & 0 & \mu_1 & (-1)^{i-1}\mu_2 \\
0 & 0 & (-1)^{i-1}\mu_2 & \mu_1
\end{array}\right],
\end{align}
where $ \mu_1 = \bar{n} + \frac{1}{2} $, $ \mu_2 = \sqrt{\bar{n}(\bar{n}+1)} $, and $ i \in \{0,1\} $.  

BPSK phase modulation of TMSV states has been shown to approach the fundamental limit of entanglement-assisted channel capacity and to significantly outperform the (unassisted) Holevo capacity of classical communication, with the capacity enhancement factor going to infinity as $ \bar{n} \to 0 $~\cite{saikat2020ISIT}. The output is pulse position modulated with the probability distribution given in Eqs.~(7) and (8) of \cite{saikat2020ISIT}, and we numerically evaluate both the relative entropy and channel capacity numerically for this setup, using system parameters $M = 2$, $K = 2$, and $n = 8$ as defined in \cite{saikat2020ISIT}.
As shown in 
Fig.~\ref{fig:RE vs Capacity},
employing TMSV input states received by two PNR detectors and an optimized squeezing parameter simultaneously improves communication capacity and decreases the change-detection latency.

Figure~\ref{fig:CUSUM_simulation} presents the numerical simulations of CUSUM test latency for three transceivers: a BPSK-modulated TMSV input state with optimized TMS and two PNR detectors, a BPSK-modulated coherent state with a homodyne detector, and a Gaussian-modulated coherent state with a homodyne detector. In each simulation, after the change occurs at time $n_c$, the data sample $x_k$ is generated from the postchange probability distribution function (PDF) $p_1$. The CUSUM statistic is updated recursively as $f_i = \max\{f_{i-1}+ \ln\frac{p_1(x_i)}{p_0(x_i)},0\}$, and a change is declared when $f_i \ge \ln(\gamma)$ at some time $n_d$, where $\gamma$ is related to the false alarm rate and is predetermined.  The detection latency is then evaluated as $t= n_d-n_c$. Each point in Fig.~\ref{fig:CUSUM_simulation} is an average of 20000 independent simulations.
As predicted by the relative entropy in 
Fig.~\ref{fig:RE vs Capacity}
the performance of the two coherent-state transceivers is close, while the BPSK-modulated TMSV transceiver has much smaller latency. 
\section{Conclusion} \label{sec: conclusion}
We analyzed the performance of various quantum transceivers in the low-noise regime. Our results revealed that the QREs for both TMSV- and coherent-state inputs scale as $ -\ln\Bar{n}_{\rm B} $, indicating that transmittance changes cannot be detected instantaneously. However, leveraging a TMSV state with a TMS and PNR detection at the receiver significantly enhances change detection. Furthermore, we prove that this approach asymptotically attains the fundamental QRE limit as $ \Bar{n} \to \infty $. Notably, the presence of a TMS at the receiver substantially increases the relative entropy when the input energy is high.  

Beyond change detection, we studied the joint communication and change-detection problem. Our findings show that a BPSK-modulated preshared TMSV state outperforms a squeezed coherent-state transceiver in terms of both channel capacity and change detection latency, particularly under low-quality channel conditions.  

\section*{Acknowledgment}
This research was supported by the DARPA Quantum Augmented Networks (QuANET) program. The views, opinions, and findings expressed in this paper are those of the authors and do not necessarily reflect the official views or policies of the Department of Defense or the U.S. Government.


\bibliography{papers.bib}

\appendix
\section{QRE of TMSV and coherent states}\label{app:QRE}
In this appendix, we derive the QRE between $m$-mode Gaussian states $\hat{\sigma}_{1}$ and $\hat{\sigma}_{0}$ by employing the Eq.~(9) in \cite{PhysRevLett.119.120501}. Let $ \mu_{\hat{\sigma}_{s}} $ and  $ \Sigma_{\hat{\sigma}_{s} }$ be the mean vector and covariance matrix of the state $\hat{\sigma}_{s}$, respectively, and $s = 0,1$. The QRE is given by
\begin{align}
    D\left(\hat{\sigma}_{1}|| \hat{\sigma}_{0} \right) &= \left[ \ln\left( \frac{Z_{\hat{\sigma}_{0}}}{Z_{\hat{\sigma}_{1}}} \right) - \trace\left\{ \Gamma \Sigma_{\hat{\sigma}_{1}} \right\} + \gamma G_{\hat{\sigma}_{0}}\gamma^T\right]/2, \nonumber \label{eq:QRE formula1}
\end{align}
where $ Z_{\hat{\sigma}_{s}}  = \det\left(\Sigma_{\hat{\sigma}_{s}}+i\Omega/2\right) $, $\Gamma  = G_{\hat{\sigma}_{1}}  - G_{\hat{\sigma}_{0}} $, $ G_{\hat{\sigma}_{s}} =  2i\Omega\arccoth\left( 2i \Sigma_{\hat{\sigma}_{s}}\Omega \right)$, $I_m $ is $m \times m$ an identity matrix, $\Omega = \begin{bmatrix}
       0 & I_{m} \\
       -I_{m} & 0
   \end{bmatrix} $, and $\gamma = \mu_{\hat{\sigma}_{1}} - \mu_{\hat{\sigma}_{0}} $.

\subsection{QRE for the TMSV input state}\label{sec:TMSV}

The TMSV state is a zero-mean pure Gaussian state characterized by the covariance matrix (in $\hat{q}\hat{q}\hat{p}\hat{p}$ form)
\begin{align}
\Sigma_{\hat{\rho}_{\mathrm{T},s}}=\left[\begin{array}{cccc}
\mu_1 & \mu_2 & 0 & 0 \\
\mu_2& \mu_1 & 0 & 0 \\
0 & 0 & \mu_1 & -\mu_2 \\
0 & 0 & -\mu_2 & \mu_1
\end{array}\right],
\end{align}
where $\mu_1 = \bar{n}+\frac{1}{2}$ and $\mu_2 = \sqrt{\bar{n}(\bar{n}+1)}$.
The signal mode of the TMSV state is sent through the lossy thermal noise bosonic channel $\mathcal{E}^{\bar{n}_{\rm B},\eta_s}$, while the idler  mode is preshared with the receiver.
The output state $\hat{\sigma}_{\mathrm{T},s} $ remains a zero-mean Gaussian state with a covariance matrix given by
\begin{align}
\Sigma_{\hat{\sigma}_{\mathrm{T},s}}& =  X\Sigma_{\hat{\rho}_{\mathrm{T},s}}X^T +Y \\
&=\left[\begin{array}{cccc}
w_{11}(s) & w_{12}(s) & 0 & 0 \\
w_{12}(s) & w_{22} & 0 & 0 \\
0 & 0 & w_{11}(s) & -w_{12}(s) \\
0 & 0 & -w_{12}(s) & w_{22}
\end{array}\right],
\end{align}
where $ X = \diag\left(\sqrt{\eta_s},1,\sqrt{\eta_s},1\right)$, 
 $Y=\diag\left( (\bar{n}_{\rm B}+\frac{1}{2})(1-\eta_s),0,(\bar{n}_{\rm B }+\frac{1}{2})(1-\eta_s),0 \right)$, and
\begin{align}
\label{eq:Appw11}	w_{11}(s) & =\bar{n}_{\rm B}(1-\eta_s)+\eta_s\bar{n}+\frac{1}{2},\\
\label{eq:Appw22}	w_{22} & =\bar{n}+\frac{1}{2}, \,{\text{and}}\\
\label{eq:Appw12}	w_{12}(s) & = \sqrt{\eta_s \bar{n} (\bar{n}+1)}.
\end{align}
The covariance matrix $\Sigma_{\hat{\sigma}_{\mathrm{T},s}}$ can be diagonalized using a two-mode squeezer (TMS) represented by a symplectic matrix:
\begin{align}
    L_{\hat{\sigma}_{\mathrm{T},s}} = \left[\begin{array}{cccc}
\sqrt{\nu_s^2+1} & \nu_s & 0 & 0 \\
\nu_s & \sqrt{\nu_s^2+1} & 0 & 0 \\
0 & 0 & \sqrt{\nu_s^2+1} & -\nu_s \\
0 & 0 & -\nu_s & \sqrt{\nu_s^2+1}
\end{array}\right],
\end{align}
where the coefficient is
\begin{align}
    \nu_{s} = \frac{1}{\sqrt{2}}\sqrt{\frac{w_{11}(s)+w_{22}}{\sqrt{(w_{11}(s)+w_{22})^2 - 4w_{12}^2(s)}}-1}.\label{eq:squeezing parameter}
\end{align}
Thus, the covariance matrix of the output state yields $ \Sigma_{\hat{\sigma}_{\mathrm{T},s}} = L_{\hat{\sigma}_{\mathrm{T},s} }\Sigma_{\hat{\sigma}^{(d)}_{\mathrm{T},s}}L^T_{\hat{\sigma}_{\mathrm{T},s} }$, where the diagonalized covariance matrix is $ \Sigma_{\hat{\sigma}^{(d)}_{\mathrm{T},s}}  = \diag\left( \bar{n}_{\mathrm{T},1}(s)+\frac{1}{2},\bar{n}_{\mathrm{T},2}(s)+\frac{1}{2},\bar{n}_{\mathrm{T},1}(s)+\frac{1}{2},\bar{n}_{\mathrm{T},2}(s)+\frac{1}{2} \right)$  and  $L_{\hat{\sigma}_{\mathrm{T},s} }^T$ is the transpose of $L_{\hat{\sigma}_{\mathrm{T},s} }$. The mean photon number of the signal and the idler mode of the two-mode thermal state $\hat{\sigma}_{\mathrm{T},s}^{(d)}$ are
\begin{align}
   \bar{n}_{\mathrm{T},1}(s)& = \frac{1}{2}\left(a(s)+w_{11}(s)-w_{22}-1\right)\label{eq:n1s}\\
  \bar{n}_{\mathrm{T},2}(s)& = \frac{1}{2}\left(a(s)+w_{22}-w_{11}(s)-1\right),\label{eq:n2s}
\end{align} 
with $  a(s)= \sqrt{(w_{11}(s)+w_{22})^2-4w_{12}^2(s) }$.
If we apply the formula \eqref{eq:QRE formula1}, the QRE between $\hat{\sigma}_{\mathrm{T},1}$ and $\hat{\sigma}_{\mathrm{T},0}$ yields
\begin{align}
    D_{\rm TMSV} 
    &= \left(b\left(1+\bar{n}_{\mathrm{T},2}(1)\right)+c\bar{n}_{\mathrm{T},1}(1)\right)^2\ln\left(1+\frac{1}{\bar{n}_{\mathrm{T},1}(0)}\right)\nonumber\\
    &\quad+\left(b\left(1+\bar{n}_{\mathrm{T},1}(1)\right)+c\bar{n}_{\mathrm{T},2}(1)\right)^2\ln\left(1+\frac{1}{\bar{n}_{\mathrm{T},2}(0)}\right) \nonumber\\
    &\quad+\ln\left( (1+\bar{n}_{\mathrm{T},1}(0))(1+\bar{n}_{\mathrm{T},2}(0))  \right) \nonumber\\
    &\quad-g(\bar{n}_{\mathrm{T},1}(1))-g(\bar{n}_{\mathrm{T},2}(1)),\label{eq:QRE TMSV}
\end{align}
where $b =\left( \nu_1\sqrt{1+\nu_0^2} - \nu_0\sqrt{1+\nu_1^2} \right)^2 $, $c = \left( \nu_0\nu_1 - \sqrt{1+\nu_1^2}\sqrt{1+\nu_0^2}  \right)^2 $, and $ g(x) = (1+x)\ln(1+x) -x\ln(x)$.
\subsection{QRE for coherent input state}
A coherent-state input with mean photon number $\bar{n}$ is characterized by the mean vector $ \mu_{\hat{\rho}_{\rm coh}}  = [\sqrt{2\bar{n}},0]$  and covariance matrix
$   \Sigma_{\hat{\rho}_{\mathrm{coh}}}= I_2/2.$
After passing through the channel, the output state has mean vector and covariance matrix $ \mu_{\hat{\sigma}_{\mathrm{coh},s}} = [\sqrt{2\eta_s\bar{n}},0]$ and $
  \Sigma_{\hat{\sigma}_{\mathrm{coh},s}} =   \left(\frac{1}{2} +\bar{n}_{\rm B}(1-\eta_s)\right) I_2 $, respectively.
The QRE between the postchange state $ \hat{\sigma}_{\mathrm{coh},1 } $  and  prechange state $ \hat{\sigma}_{\mathrm{coh},0 } $ is
\begin{align}
    &D_{\rm coh} = \ln\left(1+\bar{n}_{\rm B}(1-\eta_0) \right) \bar{n}_{\rm B}(1-\eta_1)\ln\left( \bar{n}_{\rm B}(1-\eta_1) \right) \nonumber\\
    & \quad- (1+\bar{n}_{\rm B}(1-\eta_1)) - \ln\left(1-\frac{1}{1+\bar{n}_{\rm B}(1-\eta_0)}\right) \nonumber\\
    & \quad\times \left(\bar{n}_{\rm B}(1-\eta_1)+ \bar{n}\left( \sqrt{\eta_0}-\sqrt{\eta_1} \right)^2\right).\label{eq:QRE coherent}
\end{align}

\section{Relative entropy of various transceivers}\label{sec:RE}
\subsection{Coherent-state input with a homodyne-detection receiver}
When a coherent state is transmitted through $\mathcal{E}^{\Bar{n}_{\rm B},\eta_s} $ and $ \hat{q}$ quadrature of the output state is measured via homodyne detection, the channel reduces the quantum channel to a classical lossy additive white Gaussian noise channel. The resulting data are described by  Gaussian-distributed random variables $X_{s}$, where $ X_{s}\sim\mathcal{N}\left( \sqrt{\eta_s\Bar{n}},\frac{1+2\bar{n}_{\rm B}(1-\eta_s)}{4} \right) $. The relative entropy between the postchange PDF of $X_{1}$ and prechange PDF of $X_{0} $ is 
\begin{align}
    S_{\rm coh,hom} &= \frac{\bar{n}_{\rm B}\left(\eta_0-\eta_1\right)+2\bar{n}(\sqrt{\eta_0}-\sqrt{\eta_1})^2}{1+2\bar{n}_{\rm B}(1-\eta_0)} \nonumber\\
    &\quad+ \frac{1}{2}\ln\left(\frac{1+2\bar{n}_{\rm B}(1-\eta_0)}{1+2\bar{n}_{\rm B}(1-\eta_1)}\right).\label{eq:RE coh hom}
\end{align}
\subsection{Coherent-state input with Kennedy receiver}
The output state has mean vector and covariance $[\sqrt{2\eta_s \bar{n}},0]$ and $ \left(\frac{1}{2}+\bar{n}_{\rm B}(1-\eta_s)\right)I_{2} $, respectively. The output prechange PDF is
\begin{align}
    p\left(X_0 = k\right) = \frac{(\bar{n}_{\rm B}(1-\eta_0))^k}{(1+\bar{n}_{\rm B}(1-\eta_0))^{k+1}}.
\end{align}
The output postchange PDF is
\begin{align}
    &p\left(X_1 = k\right) =e^{-\alpha_d^2} \nonumber\sum_{l=0}^{\infty}\frac{(\bar{n}_{\rm B}(1-\eta_1))^l}{(1+\bar{n}_{\rm B}(1-\eta_1))^{l+1}}\\
    &\quad\times
    \begin{cases}
     \sqrt{\frac{l!}{k!}} (-\alpha_d)^{k-l}   L_l^{(k-l)}(\alpha_d^2 ) \ \ k\ge l\\
     \sqrt{\frac{k!}{l!}} (\alpha_d)^{l-k} L_l^{(l-k)}(\alpha_d^2 ) \ \ k< l
    \end{cases},
\end{align}
where $\alpha_d = \sqrt{\bar{n}}(\sqrt{\eta_0}-\sqrt{\eta_1})$.

\subsection{Entanglement-augmented and squeezed coherent state with a homodyne detector }
An $m$-mode entanglement-augmented (EA) coherent state $\hat{\rho}_{\rm ec}$~ \cite{SaikatEAcoh} generated by passing $\ket{0,r}\otimes \ket{0}^{\otimes (m-1)}$ through the {\em Green Machine} \cite{cui2023superadditive}, which is modeled by a passive Hadamard unitary \cite{deutsch1985quantum, PhysRevLett.106.240502}, followed by applying displacement $\hat{D}\left( \alpha_r \right)$ to every $m$ mode. Here, $ \ket{0,r}$ represents a single-mode squeezed vacuum state with squeezing parameter $r$. Each mode satisfies the energy constrain such that 
$\alpha_r^2 + \frac{\sinh^2 r}{m} = \Bar{n}$.
Measuring the output state using $m$ homodyne detectors produces a multivariate Gaussian-distributed random variable $\bm{X}_{s} $ with mean vector and covariance matrix 
\begin{align}
    \mu_{\bm{X}_{s}} &= \sqrt{\eta_s}\alpha_r \mathbbm{1}_{1\times m} \label{eq:EA coherent mean} \\
    \Sigma_{\bm{X}_{s}} & = \frac{1+2\bar{n}_{\rm B}(1-\eta_s)}{4}I_{m}+\frac{\eta_s(e^{-2r}-1)}{4m}\mathbbm{1}_{m\times m},\label{eq:EA coherent variance}
\end{align}
where we define $\mathbbm{1}_{k\times l}$ as a $k$ by $l$ matrix in which all entries are ones and $I_{m}$ is an $m$ by $m$ identity matrix. We calculate the classical relative entropy of this receiver using 
\begin{align}
    S\left( \bm{X}_{1}||\bm{X}_{0} \right) &= \frac{1}{2}\left( \trace\left( \Sigma_{\bm{X}_{0}}^{-1}\Sigma_{\bm{X}_{1}} \right) - \ln \frac{|\Sigma_{\bm{X}_{1}}|}{|\Sigma_{\bm{X}_{0}}|} -m \right.\nonumber\\
    &\left.\quad+ \left(\mu_{\bm{X}_{1}}-\mu_{\bm{X}_{0}}\right)\Sigma_{\bm{X}_{1}}^{-1} \left(\mu_{\bm{X}_{1}}-\mu_{\bm{X}_{0}}\right)^T\right),
\end{align}
where $|\Sigma_{\bm{X}_{s}}|$ denotes the determinant of $\Sigma_{\bm{X}_{s}}$. The relative entropy per mode is 
\begin{align}
    S_{\rm ec,hom}(m) = \frac{S\left( \bm{X}_{1}||\bm{X}_{0} \right)}{m}. \label{eq:RE ent coh hom}
\end{align}
When $m = 1$, the EA state reduces to a squeezed coherent state, and the relative entropy satisfies
\begin{align}
    S_{\rm sc, hom} = S_{\rm ec,hom}(1).
\end{align}


\subsection{TMSV state with a TMS and PNR detectors (including SPD)}\label{sec:tmsv single pnr}
A PNR detector performs an ideal photon-number-basis measurement of the quantum state of a mode. As introduced in Sec.~\ref{sec:TMSV}, the first mode of the TMSV pass through $\mathcal{E}^{\Bar{n}_{\rm B},\eta_s}$ results in a two-mode squeezed thermal state $\hat{\sigma}_{\mathrm{T},s} = \hat{S}^{\dagger}(r_s) \hat{\sigma}_{\mathrm{T},s}^{d} \hat{S}(r_s)$, where $ \hat{S}(r_s)$ represents a TMS with squeezing parameter $r_s = -\arcsinh \nu_s $ and $\nu_s$ is defined in \eqref{eq:squeezing parameter}. In Sec.~\ref{sec:QRE of TMSV and coh}, we showed that a receiver constructed by a TMS with squeezing parameter $-r_0$ followed by a single-mode photon detector at the first mode and tracing out the second mode achieves infinite relative entropy when $\Bar{n}_{\rm B}= 0$. Now, we consider $\Bar{n}_{\rm B}>0$, the data have PDF  $ p(X_s = k) = \frac{q_s^k}{(1+q_s)^{1+k}} $, where 
\begin{align}
    q_0 &= \Bar{n}_{\mathrm{T},1}(0) \label{eq:q0}\\
    q_1 &= \Bar{n}_{\mathrm{T},1}(1)+\nu^2(\Bar{n}_{\mathrm{T},1}(1)+\Bar{n}_{\mathrm{T},2}(1)).\label{eq:q1}
\end{align}
In practice, the PNR detector has a finite photon-number-resolution limit. Suppose the resolution limit is $l$, the relative entropy is 
\begin{align}
    S_{\rm TMSV,PNR}^{(l)} = \left(1-\left( \frac{q_1}{1+q_1} \right)^{l+1}\right) S_{\rm TMSV,PNR}^{(\infty)},
\end{align}
where $S_{\rm TMSV,PNR}^{(\infty)}$ is the relative entropy using the perfect PNR detector and is given by
\begin{align}
    S_{\rm TMSV,PNR}^{(\infty)} = q_1\ln\left(\frac{q_1}{q_0}\right)+(1+q_1)\ln\left( \frac{1+q_0}{1+q_1} \right). \label{eq:RE TMSV PNR}
\end{align}
Specifically, $S_{\rm TMSV,PNR}^{(1)} $ corresponds to the case of a single-photon detector, in which the relative entropy is
\begin{align}
     S_{\rm TMSV,SPD} = \frac{q_1}{1+q_1}\ln\left(\frac{q_1}{q_0}\right)+\ln\left(\frac{1+q_0}{1+q_1}\right). \label{eq:RE TMSV SPD}
\end{align}

\subsection{TMSV input state with a TMS and a two-PNR-based receiver}
Instead of tracing out the second mode as in Sec.~\ref{sec:tmsv single pnr}, measuring the second mode with a PNR detector generates data that are described by a pair of random variables $\bm{X}_s = \left\{ X_s^{(1)},X_s^{(2)} \right\}$. The prechange PMF is
    $p_{\bm{X}_0}(k,l) = \zeta_0\left(k,l\right)$,
where we define $    \zeta_s\left(k,l\right) = \frac{\Bar{n}_{\rm T,1}(s)^k\Bar{n}_{\rm T,2}(s)^{l}}{(1+\Bar{n}_{\rm T,1}(s))^{k+1}(1+\Bar{n}_{\rm T,2}(s))^{l+1}}$. The postchange PMF is (see Eq.~(35) in \cite{zihao2023})
\begin{align}
&p_{\bm{X}_1}(k,l)  \nonumber \\
    &= \sum_{s=\max(k-l,0)}^{\infty} \zeta_1\left(s,s-k+l\right)  \tau_{1}^{2(s-k)} s!(s-k+l)! k!l!\nonumber\\
	&\times  \bigg|\sum_{u = \max(0,k-s)}^{\min(k,l)} \frac{(-\tau_1^{2})^u \tau_2^{-(k+l-2u+1)}}{(s-k+u)!u!(k-u)!(l-u)!} \bigg|^{2},  \label{eq:likelihood function}
\end{align}
where $ \tau_1 = \tanh(r_1-r_0)$ and $\tau_2 = \cosh(r_1-r_0)$.
Due to the complexity of $ p_{\bm{X}_1}(k,l) $, there is no closed form of the relative entropy $S_{\rm TMSV,PNRs}$. Therefore, we evaluate it numerically.

\subsection{TMSV input state with a TMS and a homodyne-based receiver}
Replacing the PNR detector in Sec.~\ref{sec:tmsv single pnr} with a homodyne detector generates data that are  Gaussian distributed
$X_s\sim\mathcal{N}\left(0,q_s\right)$. The relative entropy is 
\begin{align}
    S_{\rm TMSV,hom} &= \frac{(\bar{n}_{\rm B} - \bar{n})(\eta_0-\eta_1)}{1+2\bar{n}_{\rm B}(1-\eta_0)+2\bar{n}\eta_0} \nonumber \\
    &\quad+ \ln\left(\frac{1+2\bar{n}_{\rm B}(1-\eta_0) + 2\bar{n}\eta_0}{1+2\bar{n}_{\rm B}(1-\eta_0) + 2\bar{n}\eta_0}\right).
\end{align}
\subsection{Fock-state input with PNR detection at the receiver}
The Fock state is optimal in transmittance sensing in a pure-loss channel because it achieves the quantum Fisher information \cite{weedbrook12gaussianQIrmp}. The output PMF is ( see Eq.~(7.37) \cite{guha04mastersthesis})
\begin{align}
\nonumber p_X(x)&=\binom{x+\bar{n}}{x}\frac{(1-\eta_s)^{x+\bar{n}} (1+\bar{n}_{\rm B})^{\bar{n}} \bar{n}_{\rm B}^x }{ (1+\bar{n}_{\rm B}(1-\eta_s) )^{x+\bar{n}+1} }\\
\label{eq:pnk_2F1trans}&\phantom{=}\times{}_2F_1\left[\begin{array}{c}-x,-\bar{n}\\-(x+\bar{n})\end{array};z(\eta_s,\bar{n}_{\rm B})\right],
\end{align}
where $z(\eta_s,\bar{n}_{\rm B})=\frac{\left(\bar{n}_{\rm B} -\eta_s\left(1+\bar{n}_{\rm B}\right) \right) \left(1+(1-\eta_s)\bar{n}_{\rm B}\right)}{(1-\eta_s)^2\bar{n}_{\rm B}(1+\bar{n}_{\rm B})}$ and 
\begin{align}
\label{eq:2F1}{}_2F_1\left[\begin{array}{c}a,b\\ c\end{array};z\right]&=1+\frac{ab}{c}z+\frac{a(a+1)b(b+1)}{c(c+1)2!}z^2+\cdots
\end{align}
is the hypergeometric series.
We are able to calculate the relative entropy  $S_{\rm Fock,PNR}$ numerically.

\end{document}